\documentclass[11pt]{article}
\pdfoutput=1
\usepackage{jcappub,amsmath,amssymb}
\usepackage{tikz,graphics,pgfplots}
\newcommand{\St}{{St\"uckelberg} }
\def\be{\begin{equation}}
\def\ee{\end{equation}}
\def\bea{\begin{eqnarray}}
\def\eea{\end{eqnarray}}
\def\la{\langle}
\def\ra{\rangle}
\def\nn{\nonumber}
\DeclareMathOperator{\Tr}{Tr}

\begin{document}

\title{On strong coupling scales in \\(minimal) massive gravity}

\author[a]{James Bonifacio,}
\author[b]{Johannes Noller}
\affiliation[a]{Theoretical Physics, University of Oxford, DWB, Keble Road, Oxford, OX1 3NP, UK}
\affiliation[b]{Astrophysics, University of Oxford, DWB, Keble Road, Oxford, OX1 3RH, UK} 
\emailAdd{james.bonifacio@physics.ox.ac.uk}
\emailAdd{noller@physics.ox.ac.uk}

\abstract{
Ghost-free massive gravity models generically have a strong coupling scale of $\Lambda_3 =(M_{\rm Pl} m^2)^{1/3}$. However, for one of these models -- `minimal massive gravity' -- it is not clear what this scale is in the subset of solutions with vanishing vector modes, since there are then {\it no} interactions at the scale $\Lambda_3$. We show that there are always scalar-tensor interactions at a scale arbitrarily close to $\Lambda_3$ around the Minkowski vacuum solution. This explicitly confirms and completes previous research establishing that $\Lambda_3$ effectively is the maximal strong coupling scale for {\it all} ghost-free (dRGT) massive gravity models (on Minkowski).
In the process, we also revisit and clarify the construction of generic Lorentz-invariant massive gravity models, explicitly constructing an improved non-redundant expansion for these models.}


\maketitle

\section{Introduction} \label{sec-intro}

There has been much interest recently in non-linear theories of massive spin-2 fields. This is centred around the dRGT (de Rham, Gabadadze, Tolley) construction of a two-parameter family of massive gravity theories \cite{deRham:2010ik,deRham:2010kj}, which generalises the linearised Fierz-Pauli theory \cite{Fierz:1939ix} and have been shown to be ghost-free to all orders \cite{Hassan:2011hr, deRham:2011rn, Mirbabayi:2011aa, deRham:2011qq, Hassan:2011ea, Hassan:2012qv}. See \cite{Hinterbichler:2011tt, deRham:2014zqa} for reviews of massive gravity. In this paper we revisit the general massive gravity construction of \cite{deRham:2010ik} and strong coupling scales in generic massive gravity models. This scale is of primary importance, typically controlling both the regime of validity of the effective field theory in question as well as the classical (Vainshtein) scale where non-linear interactions become important. 

The original dRGT massive gravity construction made use of the \St formalism in the effective field theory formalism for massive gravitons \cite{ArkaniHamed:2002sp} (see also \cite{ArkaniHamed:2001ca,ArkaniHamed:2003vb,Schwartz:2003vj}) to identify the tuning needed to raise the low-energy cutoff of the theory, which also automatically takes care of the Boulware-Deser ghost \cite{Boulware:1973my}. 
Also see \cite{Comelli:2012vz,Kluson:2012wf,Deser:2014hga} for related constraint analyses.
This family of theories, re-summed in \cite{deRham:2010kj}, generically contains interaction terms suppressed by the lowest energy scale $\Lambda_3 =(M_P m^2)^{1/3}$ (see e.g. \cite{deRham:2010ik}, \footnote{Note that this was already shown to be the maximal strong coupling scale of massive gravity in \cite{ArkaniHamed:2002sp}, where the presence of vector modes (which do not need to be present classically) was implicitly assumed.}). However, for a particular set of parameter values, the so-called `minimal model' \cite{deRham:2010ik,Hassan:2011vm}, the scalar-tensor interactions in the $\Lambda_3$ decoupling limit all vanish, so it is unclear what is the strong coupling scale of this theory \cite{deRham:2010ik,Hinterbichler:2011tt} (ignoring vector terms---an assumption whose validity we discuss below). 

The main purpose of this paper is to investigate the nature of interactions in the minimal model in particular. In \cite{Renaux-Petel:2014pja} this was already probed for specific configurations (e.g. static spherically symmetric ones). We here complement and extend this work by investigating interaction terms in the minimal model in their general form at the level of the action without imposing any particular such configurations. We also revisit and clarify the construction of generic Lorentz-invariant massive gravity models, their strong coupling scales and regimes of validity by explicitly presenting and discussing an improved expansion based on the work of \cite{Hassan:2011vm}. 

The outline is as follows: after reviewing the setup for generic models of massive gravity in Section \ref{sec:setup}, we discuss and clarify the structure of the effective field theory expansion of the dRGT massive gravity potential at higher orders in Section \ref{sec:expansions}. We point out an effective field theory expansion without redundancies, and we explicitly write down the coefficients in this expansion up to sixth order. In Section \ref{sec:minimal}, we use this result and the re-summed theory to show that the minimal model of massive gravity contains scalar-tensor interactions suppressed by energy scales arbitrarily close to $\Lambda_3$. We conclude in Section \ref{sec-conc}.

\section{The setup: A generic massive gravity theory}
\label{sec:setup}

Adding a Lorentz-invariant mass term (i.e. any non-derivative, potential-like self-interaction) to Einstein gravity requires introducing a non-dynamical absolute metric \footnote{Unless we forgo locality \cite{Jaccard:2013gla}.}, $g^{(0)}_{\mu \nu}$, since no potential interactions other than a cosmological constant can be built using only a metric and its inverse. Since $g^{(0)}_{\mu \nu}$ does not transform as a rank-2 tensor, this explicitly breaks diffeomorphism invariance of the action. 
Here we revisit the construction of Lorentz-invariant massive gravity theories with general potential interaction terms. These interactions will also be the most relevant ones at low energies when compared to higher order derivative interactions. Adding a general potential then gives the action
\begin{equation}
\label{eq:genericaction}
S= \frac{M_{p}^2}{2} \int d^4 x\ \left[\sqrt{-g}R-\sqrt{-g}\frac{1}{4}m^2V(g,h)\right],
\end{equation}
where $h_{\mu \nu} = g_{\mu \nu} - g^{(0)}_{\mu \nu}$ is the (massive) spin-2 field. 
For the rest of this paper we consider a flat non-dynamical reference metric, $g^{(0)}_{\mu \nu} = \eta_{\mu\nu}$, although dRGT massive gravity can be generalised to general reference metrics \cite{Hassan:2011tf} (see, e.g. \cite{deRham:2012kf} for a de Sitter reference metric).

\subsection{The potential}

A general potential can be expanded order-by-order as
\be
\label{eq:expansion}
V(g,h)=V_2(g,h)+V_3(g,h)+V_4(g,h)+V_5(g,h)+\cdots,
\ee
where the individual orders are given by \footnote{Starting the expansion at quadratic order ensures that Minkowski is a solution for the full metric.}
\begin{subequations}
\label{eq:potentialexpansion}
\bea
V_2(g,h)&=& b_1 \la h^2\ra + b_2 \la h\ra^2, \\
V_3(g,h)&=& c_1\la h^3\ra+c_2 \la h^2\ra \la h\ra +c_3\la h \ra ^3, \\
V_4(g,h)&=& d_1\la h^4\ra+d_2 \la h^3\ra \la h\ra +d_3\la h^2 \ra^2 +d_4\la h^2\ra\la h\ra^2+d_5\la h\ra^4,\\
V_5(g,h)&=& f_1\la h^5\ra+f_2 \la h^4\ra \la h\ra +f_3\la h^3 \ra \la h\ra^2 +f_4\la h^3\ra\la h^2\ra+f_5\la h^2\ra^2\la h\ra + f_6\la h^2\ra\la h\ra^3+\nn \\ &&f_7\la h\ra^5, \\
V_6(g,h)&=& g_1 \la h^6 \ra+ g_2\la h^5 \ra \la h \ra + g_3 \la h^4 \ra \la h^2 \ra+ g_4\la h^4 \ra\la h^1 \ra^2+ g_5\la h^3 \ra^2+ g_6 \la h^3 \ra\la h^2 \ra\la h^1 \ra+ \nn \\ && g_7 \la h^3 \ra\la h \ra^3+ g_8\la h^2 \ra^3+ g_9\la h^2 \ra^2\la h \ra^2+ g_{10}\la h^2 \ra\la h \ra^4+ g_{11} \la h \ra^6, \\
&\vdots& \nn
\eea
\end{subequations}
where angled brackets denote tracing with the dynamical metric, e.g. 
\begin{equation}
\label{eq:metriclowering}
\la h^3\ra = h_{\mu}^{\; \; \alpha} h_{\alpha}^{\; \; \nu} h_{\nu}^{\; \; \mu} = g^{\beta \alpha} g^{\gamma \nu} g^{\mu \sigma} h_{\mu \beta} h_{\alpha \gamma} h_{\nu \sigma}.
\end{equation}
At each order, $n$, the number of terms in the above expansion of the potential is equal to the number of distinct integer partitions of $n$, $p(n)$, so that it appears as if each potential is defined by an infinite number of parameters, $\sum_n p(n)$. As we note later, there is a redundancy in this parameterisation and consequently the number of independent terms is less than this representation implies, although still infinite. 

The Fierz-Pauli theory \cite{Fierz:1939ix} is given by the lowest (quadratic) order part of \eqref{eq:genericaction} with the parameter choice $b_2=-b_1$. This tuning is required to ensure that at lowest order, i.e. in the linear theory, there is no propagating ghost degree of freedom around a flat background.

\subsection{The \St trick and degrees of freedom}
\label{Stuk}
The \St trick \cite{stuckelberg, Siegel:1993sk,ArkaniHamed:2002sp} allows one to re-introduce diffeomorphism invariance into the action at the expense of adding additional fields. In particular, we add four fields, $Y^A$, in the combination
\begin{equation}
\label{eq:streplacement}
H_{\mu \nu}(x) = g_{\mu \nu}(x) - \eta_{A B}(Y(x)) \partial_{\mu} Y^A \partial_{\nu} Y^B,
\end{equation}
and impose that the $Y^A$ transform as scalars under diffeomorphisms. The combination $H_{\mu \nu} $ then transforms as a rank-two tensor under diffeomorphisms and we can define a diffeomorphism invariant version of the action \eqref{eq:genericaction} by making the replacement $h_{\mu \nu}(x) \rightarrow H_{\mu \nu}(x)$ in the potential term \footnote{This requires first lowering all indices on $h$ using the full metric, as in \eqref{eq:metriclowering}.}. The resultant theory has the same dynamical content as the original one, as can be seen by choosing the unitary gauge $Y^A =x^A$, which eliminates the additional fields introduced by the \St trick. 
However, this replacement is useful for separating the different helicity degrees of freedom of the theory and making their interaction scales explicit. To see this, one first writes $Y^{\mu} = x^{\mu} -A^{\mu}$, where we have switched to Greek indices in anticipation of the fact that the four fields $A^{\mu}$ will, in the decoupling limit (see section \ref{sec:minimal}),
transform as the components of a Lorentz four-vector \cite{Mirbabayi:2011aa}. By further replacing $A^{\mu}$ with $A^{\mu} + \partial^{\mu} \phi$, we introduce the field $\phi$, which will transform as the helicity-0 component of the graviton in the decoupling limit. The net effect of this replacement is then to replace $h_{\mu \nu}$ in the potential term (not including $\sqrt{-g}$ or $g^{\mu \nu}$) with
\begin{equation}
\label{eq:StRep}
H_{\mu \nu} = h_{\mu \nu} + \partial_{\mu} A_{\nu} + \partial_{\nu} A_{\mu} +2\partial_{\mu} \partial_{\nu} \phi - \partial_{\mu} A^{\alpha} \partial_{\nu} A_{\alpha}-\partial_{\mu}A^{\alpha} \partial_{\nu} \partial_{\alpha} \phi - \partial_{\mu} \partial^{\alpha} \phi \partial_{\nu} A_{\alpha} - \partial_{\mu}\partial^{\alpha} \phi \partial_{\nu} \partial_{\alpha} \phi.
\end{equation}
An equivalent representation of a generic massive gravity theory, which is manifestly diffeomorphism invariant and useful for writing down the re-summed theories, is to work directly in terms of the rank-two tensor field 
\be \label{HRf}
f_{\mu \nu}(x) = \eta_{A B}(Y(x)) \partial_{\mu} Y^A \partial_{\nu} Y^B,
\ee
and to construct the potential out of scalar functions of the tensor $g^{\mu \alpha} f_{\alpha \nu}$. Note that we could have introduced the \St degrees of freedom in a different way, e.g. via \St transforming $g_{\mu\nu}$ instead of the background metric. 

\subsection{Interaction scales}
\label{subsec:int}

\begin{figure}
\begin{center}
\includegraphics{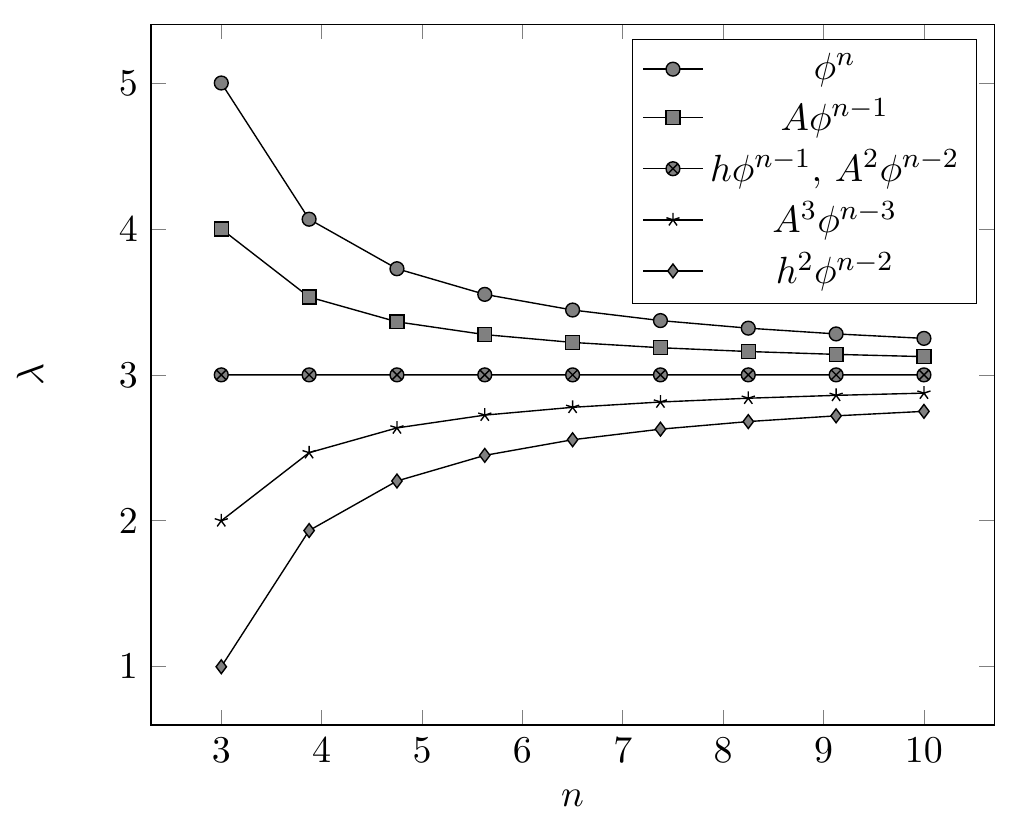} 
\caption{Scaling of $\lambda$ as given in \eqref{intscales} for different interaction terms. $n$ counts the total number of fields participating in the interaction and the different curves correspond to interaction terms as detailed in the legend. Note that a smaller $\lambda$ corresponds to a larger energy scale $\Lambda_\lambda$ and, as such, interactions suppressed by $\Lambda_5$ are suppressed by the lowest scale and hence are the most important at low energies if present.} \label{fig:scaling}
\end{center}
\end{figure}

After making the \St replacement and expanding the metric terms in the potential using $h_{\mu \nu}=g_{\mu \nu}-\eta_{\mu \nu}$, the action \eqref{eq:genericaction} consists of kinetic and potential interaction terms for the fields $h_{\mu \nu}$, $A_{\mu}$, and $\phi$. 
Canonically normalising the (linearised) kinetic terms for $h,A,\phi$ then requires us to perform the field rescalings \footnote{Technically, one should first demix kinetic terms at quadratic order (so that $h,A,\phi$ describe independent propagating degrees of freedom in the linear theory), which requires us to perform the linearised conformal transformation
\be \label{linconf}
h_{\mu\nu} \to h_{\mu\nu} + b_1 m^2 \phi \eta_{\mu\nu},
\ee
in a theory with Fierz-Pauli tuning. However, we here impose the correct normalisations, anticipating that this will give the correct kinetic structure.}:
\begin{align}
\hat{h}_{\mu \nu} &= \frac{1}{2} M_{P} h_{\mu \nu},  &\hat{A}^{\mu} &= \frac{1}{2}mM_P A^{\mu},  &\hat{\phi} &= \frac{1}{2} m^2 M_P \phi.
\end{align}
After this rescaling, and remembering the overall factor of $M_P^2 m^2$ in front of the potential in \eqref{eq:genericaction}, the general interaction term $\{n_h, n_A, n_\phi\}$ involving $n_\phi$ fields $\phi$, $n_A$ fields $A_\mu$ and $n_h$ fields $h_{\mu\nu}$ is given by \cite{Schwartz:2003vj}
\be \label{curly}
\{n_h, n_A, n_\phi\} \equiv \Lambda_{\lambda}^{4-n_h-2n_A-3n_{\phi}} \hat{h}^{n_h} (\partial \hat{A})^{n_A} (\partial^2 \hat{\phi})^{n_{\phi}},
\ee
where one can read off that the associated interaction scale is
\begin{equation} \label{intscales}
\Lambda_{\lambda}=\left(M_Pm^{\lambda-1}\right)^{1/\lambda}= m\left( \frac{M_P}{m} \right)^{1/ \lambda} , \ \ \ \lambda={3n_\phi+2n_A+n_h-4\over n_\phi+n_A+n_h-2},
\end{equation}
where $n_h + n_A + n_{\phi} \ge 3$. We emphasise that smaller $\lambda$ are associated with higher energy scales $\Lambda_\lambda$. From this, and taking $m < M_P$, it follows that the interaction suppressed by the lowest energy scale is $(\partial^2 \hat{\phi})^3$ at $\Lambda_5$, followed by $(\partial^2 \hat{\phi})^4$ and $\partial \hat{A} (\partial^2 \hat{\phi})^2$ at $\Lambda_4$. More generally, all interactions suppressed by energy scales below $\Lambda_3$ take the form $(\partial^2 \hat{\phi})^n$ or $\partial \hat{A} (\partial^2 \hat{\phi})^n$ and all interactions suppressed by $\Lambda_3$ take the form $\hat h (\partial^2 \hat{\phi})^n$ or $(\partial \hat{A})^2 (\partial^2 \hat{\phi})^n$. 
Now, for any fixed values of $n_h$ and $n_A$ such that $n_h+\frac{n_A}{2} > 1$, we have that $\Lambda_{\lambda}$ tends to $\Lambda_3$ from above as $n_{\phi}$ tends to infinity \footnote{That is, for any $\epsilon > 0$, interactions of the form $\hat{h}^{n_h} (\partial \hat{A})^{n_A} (\partial^2 \hat{\phi})^{n_{\phi}}$ are suppressed by energy scales $\Lambda_{3-\epsilon}$, for large enough $n_{\phi} \gg n_h , n_A$.}.
`From above' here means that such interactions have energy scales larger than $\Lambda_3$, corresponding to smaller $\lambda$. 
It is worth emphasising that this means that there are infinitely many interaction terms arbitrarily close to $\Lambda_3$ both from above and below. Figure \ref{fig:scaling} summarises these statements.

\section{Raising the strong coupling scale}
\label{sec:expansions}

A generic massive gravity model as parametrised in the previous section is an effective field theory with irrelevant (non-renormalizable) interaction terms suppressed by the scales $\Lambda_\lambda$. The lowest of the scales present, which depends on the choice of potential $V$, we will refer to as the strong coupling scale of the theory, $\Lambda_{\rm strong}$. Without any particular tuning of the free coefficients in the potential this scale is $\Lambda_5$ \cite{ArkaniHamed:2002sp, Aubert:2003je}. This untuned theory has several problems. 
Most important of all, it makes the theory essentially non-predictive. Since $\Lambda_5 \sim 10^{-11} \text{ km}^{-1}$ \footnote{This value is computed for a solar mass source, $M \sim 10^{30} \text{ kg}$ and a Hubble-mass graviton $m \sim 10^{-33} \text{ eV}$.}, quantum corrections are not suppressed already at scales of $\Lambda_Q \sim 10^{-24} \text{ km}^{-1}$, i.e. scales of roughly the size of the observable universe \cite{Hinterbichler:2011tt}. In fact, the $\Lambda_5$ interactions also excite a ghost at the same scale $\Lambda_Q$ \cite{Creminelli:2005qk,Deffayet:2005ys}, so the scale where quantum corrections become important (which directly derives from $\Lambda_5$ around a given background) should indeed be seen as the effective cutoff scale of the theory in question \footnote{A detailed discussion of the relation between the cutoff of the theory---the scale where full unitarity breaks down---and the perturbative unitarity violation scales, as well as their relation to an `environmental' strong coupling scale, is beyond the scope of this paper. We refer to \cite{Burrage:2012ja, deRham:2014wfa,Kaloper:2014vqa} and references therein.}. 

This means that it is imperative to raise the strong coupling scale (and in the process also the scale of any would-be Boulware-Deser ghosts). In fact, even before worrying about ghost-freedom and Vainshtein screening, the requirement of obtaining a theory that is predictive over as large a range of energies as possible instructs us to try to raise the strong coupling scale of the theory as far as possible. This is what we will do in the remainder of this paper. 

\subsection{The strong coupling scale}

Before systematically raising the strong coupling scale in a generic massive gravity theory, let us briefly clarify what we mean by a `strong coupling scale' here, since this term is used in different ways throughout the literature. The scale of the least suppressed irrelevant operators as discussed above in section \ref{subsec:int} is what we call the strong coupling scale $\Lambda_{\rm strong}$. 
Other scales of interest are the scale $\Lambda_Q$ mentioned above, where loop corrections are no longer suppressed with respect to the tree level amplitude; the scale where full unitarity is lost, i.e. the cutoff of the theory, $\Lambda_{\rm cutoff}$; and the Vainshtein scale around a given configuration (e.g. a static and spherically symmetric massive source) $\Lambda_{V}$. 

The Vainshtein scale $\Lambda_{V}$ is sensitive to the source configuration chosen and describes the scale where classical non-linearities begin to dominate over the classical linear solution in the given configuration. $\Lambda_{\rm cutoff}$ corresponds to the scale where the theory becomes ill-defined and where full (rather than just perturbative) unitarity is lost. We emphasise that in principle this is distinct from all the scales mentioned above. All the above scales, except for the cutoff scale, do, however, directly depend on the value of $\Lambda_{\rm strong}$, so in this paper we will solely focus on identifying this scale and leave a more detailed investigation of the other derivative scales to future work.

\subsection{Vanishing combinations}

From the discussion in subsection \ref{subsec:int}, it is apparent that in order to raise the strong coupling scale in the action \eqref{eq:genericaction} we need to begin by eliminating the lowest-order $\phi$ self-interactions and those involving $\phi$ and a single vector field, since these are suppressed by the lowest energy scales.
In fact, it will turn out that eliminating the former ensures that the latter are eliminated. 
This can be achieved by choosing values for the coefficients in the potential of \eqref{eq:genericaction} such that the relevant Lagrangian terms are equal to total derivatives or are identically zero. Thus, we need the most general Lorentz-invariant total derivative and (non-trivial) zero combinations that can be made algebraically from a rank-two tensor.
To this end, consider a $4 \times 4$ matrix, $M$, with eigenvalues $\lambda_i$, $i=1,\ldots,4$. We will work in four dimensions, but the discussion easily generalises. Then
\begin{align}
\det (1 + M) & =  (1+\lambda_1)(1+\lambda_2)(1+\lambda_3)(1+\lambda_4) \nonumber \\
& = \mathcal{L}_0(M) + \mathcal{L}_{1}(M) + \frac{1}{2!}\mathcal{L}_{2}(M)+\frac{1}{3!}\mathcal{L}_{3}(M) +\frac{1}{4!}\mathcal{L}_{4}(M),
\end{align}
where the second line comes from expanding and defining $\mathcal{L}_{k}(M)/k!$ to denote the sum of all distinct products of $k$ distinct eigenvalues, e.g.
\[
\frac{1}{3!}\mathcal{L}_{3}(M) = \lambda_1 \lambda_2 \lambda_3 + \lambda_1 \lambda_2 \lambda_4+\lambda_1 \lambda_3 \lambda_4+\lambda_2 \lambda_3 \lambda_4,
\]
and $\mathcal{L}_{k}(M)$ vanishes for $k>4$. Newton's identities then give relations between $\mathcal{L}_{k}(M)$ and the $k$-th power sums:
\begin{equation}
\label{eq:Newton}
\frac{\mathcal{L}_{k}(M)}{(k-1)!} = \sum_{j=1}^k (-1)^{j-1} \frac{\mathcal{L}_{k-j}(M)}{(k-j)!}\rho_j(\lambda_i),
\end{equation}
where $\rho_j(\lambda_i) = \sum_{k=1}^4 \lambda_k^j$. We can also write $\rho_j(\lambda_i) = \left[ M^m \right]$, where parentheses denote a trace and the usual matrix product is used, i.e. $M^m = M_{\mu_1}^{ \; \; \mu_2} M_{\mu_2}^{\; \; \mu_3} \ldots M_{\mu_m}^{\; \; \mu_1}$. Thus, Newton's identities give the following expressions for $\mathcal{L}_k(M)$ in terms of traces of powers of $M$ for $k=1, \ldots, 4$:
\begin{subequations}
\label{eq:Lexpanded}
\begin{align}
\mathcal{L}_0(M) &= 1 \\
\mathcal{L}_1(M)&= \left[ M \right] \\
\mathcal{L}_2(M)& =\left[ M \right]^2 - \left[ M^2 \right] \\
\mathcal{L}_3(M) & = \left[ M \right]^3 - 3 \left[ M \right]\left[ M^2 \right]+2\left[ M^3 \right] \\
\mathcal{L}_4(M) & = \left[ M \right]^4 - 6 \left[ M^2 \right]\left[ M \right]^2+8\left[ M^3 \right]\left[ M \right]+3\left[ M^2 \right]^2-6\left[ M^4 \right].
\end{align}
\end{subequations}
For $k>4$, the left-hand side of \eqref{eq:Newton} vanishes and we get expressions such as
\begin{subequations}
\begin{align}
\label{eq:L5}
\mathcal{L}_5(M) = & \left[ M \right]^5 - 10 \left[ M^2 \right]\left[ M \right]^3+15\left[ M^2 \right]^2\left[ M \right]+20\left[ M^3 \right]\left[ M \right]^2-20\left[ M^3 \right]\left[ M^2 \right] \nonumber \\ & -30\left[ M^4 \right]\left[ M \right]+24\left[ M^5 \right] \equiv 0 \\
\mathcal{L}_6(M)  =&  \left[ M \right]^6 - 15 \left[ M^2 \right]\left[ M \right]^4+45\left[ M^2 \right]^2\left[ M \right]^2-15\left[ M^2 \right]^3+40\left[ M^3 \right]\left[ M \right]^3 \nonumber \\ & -120\left[ M^3 \right]\left[ M^2 \right]\left[ M \right]+40\left[ M^3 \right]^2-90\left[ M^4 \right]\left[ M \right]^2 +90 \left[ M^4\right]\left[ M^2 \right] \nonumber \\ & +144\left[ M^5 \right]\left[ M \right]-120\left[ M^6 \right] \equiv 0,
\end{align}
\end{subequations}
which identically vanish. We can also rearrange the expressions \eqref{eq:Lexpanded} to write $\left[ M^m\right]$ as a polynomial in $\mathcal{L}_i(M)$:
\begin{subequations}
\label{eq:trace-as-TD}
\begin{align}
\left[M^2\right]&=  \mathcal{L}_1^2-\mathcal{L}_2 \\
\left[M^3\right]&= \mathcal{L}_1^3-\frac{3}{2}\mathcal{L}_1\mathcal{L}_2 +\frac{1}{2} \mathcal{L}_3\\
\left[M^4\right]&= \mathcal{L}_1^4 - 2 \mathcal{L}_1^2 \mathcal{L}_2+\frac{2}{3} \mathcal{L}_1 \mathcal{L}_3+\frac{1}{2}  \mathcal{L}_2^2-\frac{1}{6} \mathcal{L}_4.
\end{align}
\end{subequations}
However, since $\mathcal{L}_{k>4}(M) \equiv 0$, we can also use \eqref{eq:Newton} to write $\left[ M^{m>4}\right]$ as a polynomial in $\left[ M^{m\leq4}\right]$, e.g. \eqref{eq:L5} gives
\begin{align}
\label{eq:A5}
\left[ M^5 \right] = &- \frac{1}{24}\Big( \left[ M \right]^5 - 10 \left[ M^2 \right]\left[ M \right]^3+15\left[ M^2 \right]^2\left[ M \right]+20\left[ M^3 \right]\left[ M \right]^2  \nn \\
&- 20\left[ M^3 \right]\left[ M^2 \right] -30\left[ M^4 \right]\left[ M \right] \Big).
\end{align}
 Hence, using the expressions \eqref{eq:trace-as-TD}, $\left[ M^{m>4}\right]$ can be written as a polynomial in $\mathcal{L}_{i\leq 4}(M)$ \footnote{This is a particular example of the fundamental theorem of symmetric polynomials, saying that any symmetric polynomial in the $\lambda_i$'s can be written uniquely as a polynomial in (the non-vanishing) $\mathcal{L}_i(M)$. This tells us the uniqueness at each order of non-trivially vanishing expressions such as \eqref{eq:L5} and that the only identically vanishing polynomial in the variables $\mathcal{L}_{i=1,2,3,4}$ is the zero polynomial (the polynomial with all coefficients set to zero).}. 

Thus, a better expansion of the potential, which has no redundancies \footnote{There can still be terms in the re-summed action that do not contribute to the equations of motion \cite{Hassan:2011vm}.}, differs from \eqref{eq:potentialexpansion} at orders greater than four as
\begin{subequations}
\label{eq:potentialexpansion2}
\begin{align}
V_5(g,h)=& f'_1 \la h^4\ra \la h\ra +f'_2\la h^3 \ra \la h\ra^2 +f'_3\la h^3\ra\la h^2\ra+f'_4\la h^2\ra^2\la h\ra + f'_5\la h^2\ra\la h\ra^3+ f'_6\la h\ra^5 \\
V_6(g,h)=& g'_1 \la h^4 \ra \la h^2 \ra+ g'_2\la h^4 \ra\la h^1 \ra^2+ g'_3\la h^3 \ra^2+ g'_4 \la h^3 \ra\la h^2 \ra\la h^1 \ra+ \nn \\ & g'_5 \la h^3 \ra\la h \ra^3+ g'_6\la h^2 \ra^3+ g'_7\la h^2 \ra^2\la h \ra^2+ g'_{8}\la h^2 \ra\la h \ra^4+ g'_{9} \la h \ra^6,\\
&\vdots \nn
\end{align}
\end{subequations}
That this expansion contains as much information as \eqref{eq:potentialexpansion} can be seen by substituting \eqref{eq:A5} and its generalisations into \eqref{eq:potentialexpansion}. Clearly, we could also expand in polynomials in $\mathcal{L}_{i=1,2,3,4}(h)$ using \eqref{eq:trace-as-TD}. Either way, this shows that the number of algebraically independent terms at each order $n$ is the number of partitions of $n$ involving only $1, 2, 3,$ and $4$, rather than $p(n)$. For later reference, we also note that one can convert between $\la h^m \ra$ and $\mathcal{L}_i(\mathcal{K})$, where 
\[
\mathcal{K}^{\mu}_{\nu} = \delta^{\mu}_{\nu} - \sqrt{\delta^{\mu}_{\nu} - H^{\mu}_{\nu}},
\]
which is used in the resummation of dRGT massive gravity \cite{deRham:2010kj}. 
 If $H^{\mu}_{\nu}$ has eigenvalues $\lambda_i$, then $\mathcal{K}^{\mu}_{\nu}$ has eigenvalues $\bar{\lambda}_i= 1 - \sqrt{1-\lambda_i}$. So, using $\lambda_i = 2 \bar{\lambda}_i - \bar{\lambda}_i^2$, we get, for example,
 \begin{align*}
 \la H \ra  &= 2 \la \mathcal{K} \ra - \la \mathcal{K}^2 \ra.
 \end{align*}
 We obtain similar expressions at higher orders, which, after using \eqref{eq:trace-as-TD}, give the relations:  
 \begin{subequations}
 \label{eq:H-to-TDK}
 \begin{align}
 \la H \ra  &= 2 \mathcal{L}_1- \mathcal{L}_1^2+ \mathcal{L}_2 \\
 \la H^2 \ra  &= 4 \mathcal{L}_1^2-4\mathcal{L}_1^3+\mathcal{L}_1^4-4\mathcal{L}_2+6\mathcal{L}_1\mathcal{L}_2-2\mathcal{L}_1^2\mathcal{L}_2
 +\frac{\mathcal{L}_2^2}{2} -2\mathcal{L}_3+\frac{2\mathcal{L}_1\mathcal{L}_3}{3}-\frac{\mathcal{L}_4}{6},
 \end{align}
 \end{subequations} 
where all $\mathcal{L}_i = \mathcal{L}_i(\mathcal{K})$. The expressions for $\la H^3 \ra$ and $\la H^4 \ra$, which we omit, have 23 and 46 terms, respectively. 

\subsection{Total derivatives}
Now consider the matrix of derivatives $\Pi_{\mu \nu} = \partial_{\mu} \partial_{\nu} \phi$, which appears in the \St replacement \eqref{eq:streplacement}. In fact, $\mathcal{L}_i (\Pi)$ is a total derivative for $i=1, \ldots, 4$, and these are the only such total derivative combinations \cite{Creminelli:2005qk, Nicolis:2008in}. That they are total derivatives is easiest to see by noting that we can write
\begin{align}
\label{eq:TDeta}
\mathcal{L}_i(\Pi) &= \sum_p (-1)^p \eta^{\mu_1 p(\nu_1)} \eta^{\mu_2 p(\nu_2)}\ldots \eta^{\mu_i p(\nu_i)} \Pi_{\mu_1 \nu_1} \Pi_{\mu_2 \nu_2}\ldots\Pi_{\mu_i \nu_i},
\end{align}
where the sum is over all permutations of $\nu_k$ and $(-1)^p$ is the parity of the permutation. Then, by using the antisymmetry of $\mu_1 \mu_k$ on the $\eta$'s and the symmetry of  $\partial_{\mu_1} \partial_{\mu_k}$, we get
\[
\mathcal{L}_i(\Pi) = \partial_{\mu_1} \left( \sum_p (-1)^p \eta^{\mu_1 p(\nu_1)} \eta^{\mu_2 p(\nu_2)}\ldots \eta^{\mu_i p(\nu_i)} (\partial_{\nu_1} \phi) \Pi_{\mu_2 \nu_2}\ldots\Pi_{\mu_i \nu_i} \right).
\]

\subsection{The dRGT construction}

Armed with the above, one may now begin to systematically eliminate the lowest scale interactions, thus raising the strong coupling scale of the theory. Beginning with the $\phi$ self-interactions (since these are the ones suppressed by the lowest scales), it in fact turns out that {\it all} the $\phi$ self-interactions can be removed with a careful choice of the potential coefficients. To do this, one substitutes the \St expansion into the potential \eqref{eq:potentialexpansion}, isolates the $\phi$ self-interaction terms (by replacing $\la H^n \ra $ with $\la (2\Pi - \Pi^2)^n \ra$), and then chooses the coefficients such that at each order of $\Pi_{\mu \nu} = \partial_{\mu} \partial_{\nu} \phi$ these are proportional to $\mathcal{L}_{n}(\Pi)$ \cite{deRham:2010ik}.
Ensuring that the $\phi$ self-interactions arrange themselves into multiples of total derivatives up to fourth order determines the coefficients of \eqref{eq:potentialexpansion} to be
\begin{align} c_1&=2c_3+\frac 1 2, &c_2&=-3c_3-\frac 12,\end{align}
and
\begin{align}
d_1&=-6d_5+\frac{1}{16}(24c_3+5), &d_2&=8d_5-\frac{1}{4}(6c_3+1),\nn \\
d_3&=3d_5-\frac{1}{16}(12c_3+1), &d_4&=-6d_5+\frac34 c_3.
\end{align}
Setting the fifth order $\phi$ self-interactions equal to some multiple of the zero combination $\mathcal{L}_5 (\Pi)$ then determines
\begin{align}
f_1&=\frac{7}{32}+\frac{9}{8}c_3-6d_5+24f_7, \quad \quad &f_2 &= -\frac{5}{32} -\frac{15}{16}c_3+6d_5-30f_7,\nn\\
f_3&=\frac38 c_3-3d_5+20 f_7,  &f_4&=-\frac{1}{16}-\frac34 c_3+5d_5-20f_7 ,\nn\\
f_5&=\frac{3}{16} c_3-3d_5+15f_7,  &f_6&=d_5-10f_7,
\end{align}
where $f_7$ is left arbitrary. Similarly, at sixth order $g_{11}$ is left as an arbitrary overall coefficient. It appears, therefore, that by eliminating the $\phi$ self-interactions we pick up a free coefficient at every order. In fact, it is apparently even worse than this, since at higher orders there are more and more vanishing combinations available to set the $\phi$ self-interactions equal to, since we can consider lower order vanishing $\mathcal{L}_i(\Pi)$'s times some arbitrary polynomial. For example, at sixth order we can set the $\phi$ self-interactions equal to any linear combination of $\left[ \Pi \right] \mathcal{L}_5$ and $\mathcal{L}_6$, since these both vanish. This means that there is another redundant coefficient, which we can choose to be $g_{10}$.

However, dRGT, as uniquely defined by requiring the absence of pure $\phi$ self-interactions from the potential, is only a two-parameter family of theories \footnote{In addition to the graviton mass and cosmological constant, for fixed $M_P$.}, not an infinite-parameter one. This means that all the apparently free parameters actually do not appear in the action. What happens is that these identically vanishing combinations of $\Pi$ terms come about from identically vanishing combinations of $H$ terms, so the arbitrary coefficients always end up multiplying zero combinations in \eqref{eq:potentialexpansion}. For example, as pointed out in \cite{deRham:2010ik}, at fifth order $f_7$ multiplies $\mathcal{L}_5(H) \equiv 0$. Due to mixing between orders in going from $H$ to $\Pi$, $f_7$ will appear at orders five through ten in \eqref{eq:potentialexpansion}, and must always multiply a zero combination; indeed, at sixth order $f_7$ multiplies the combination $\mathcal{L}_6 - \left[ 1 \right] \mathcal{L}_5 \equiv 0$. There are, in principle, infinitely many other terms to check if we want to proceed this way, but they are guaranteed to vanish by the observation that the combinations $\mathcal{L}_i(\Pi)$ can only come about from $\mathcal{L}_i(1\pm\sqrt{1-H})$, since $1\pm\sqrt{1-H}$ are the only Lorentz-invariant functions of $H$ that reduce to $\Pi$ when $h_{\mu \nu}=A_{\mu}=0$. However, $\mathcal{L}_{i>4}$ vanishes identically, i.e. no matter what the argument is, so $V(g,h)$ is unaffected by adding $\mathcal{L}_{i > 4}(1\pm\sqrt{1-H})$. 

\subsection{dRGT in the improved expansion}

All of this means that the set of coefficients defining a specific dRGT theory in the expansion \eqref{eq:potentialexpansion} are not unique---we can always add some combination of $\mathcal{L}_{i>4}(H)$ terms that will affect these coefficient values without altering the theory. If we want to characterize our theory with some unique set of coefficients, we ought to use the parameterisation \eqref{eq:potentialexpansion2}. 
This has the advantage that the general process outlined above is now modified such that all coefficients of $\Pi$ self-interactions at orders greater than four need to be set to zero exactly, since there are no non-trivial vanishing combinations of $\mathcal{L}_{i \leq 4}(\Pi)$. This eliminates all redundancies that can appear from a finite number of terms and makes it clear that there are only two free parameters. It also gives another way of deriving the dRGT potential: substitute the expressions \eqref{eq:H-to-TDK} into the generic potential in the improved expansion and tune coefficients to leave a first-order polynomial in $\mathcal{L}_{i=2,3,4}(K)$.
In any case, the potential coefficients in \eqref{eq:potentialexpansion2} up to sixth order are the following (up to fourth order the two expansions are identical):

\begin{align}
f_1'&=\frac{15}{128} +\frac{15}{32}c_3-\frac{3}{2}d_5,  &f_2' &= -\frac{35}{192}-\frac{9}{16}c_3+2 d_5,\nn \\
f_3'&=\frac{23}{192}+\frac{3}{16}c_3,  &f_4'&=-\frac{35}{256}-\frac{33}{64}c_3+\frac{3}{4}d_5,\nn\\
f_5'&=\frac{35}{384}+\frac{15}{32}c_3-\frac{3}{2}d_5,  &f_6'&=-\frac{7}{768}-\frac{3}{64}c_3+\frac{1}{4}d_5,
\end{align}

\begin{align}
g_1'&=\frac{43}{512}+\frac{3}{16}c_3-\frac{9}{16}d_5,  & g_2' &= -\frac{7}{512}+\frac{9}{128}c_3-\frac{3}{16}d_5,\nn\\
g_3'&= \frac{5}{128}+\frac{5}{48}c_3,  &g_4'&=-\frac{35}{384}-\frac{23}{64}c_3+\frac{3}{4}d_5 ,\nn\\
g_5'&=-\frac{7}{384}-\frac{7}{192}c_3+\frac{1}{4}d_5 ,  &g_6'&=-\frac{21}{1024}-\frac{3}{32}c_3+\frac{9}{32}d_5,\nn\\
g_7'&=\frac{7}{1024}+\frac{21}{256}c_3-\frac{15}{32}d_5 , & g_8'&=\frac{49}{3072}+\frac{7}{128}c_3-\frac{3 }{32}d_5, \nn\\
g_9'&=-\frac{7}{3072}-\frac{7}{768}c_3+\frac{1}{32}d_5.
\end{align}

We can now perform the \St expansion on this form of the potential to see the interactions between the helicity fields up to some given order, without worrying about square roots. However, without the terms up to arbitrarily high orders, we can of course only study the theory perturbatively.

The improved expansion also offers a straightforward way to see that dRGT massive gravity is only a two-parameter family of theories in four dimensions. At orders greater than four, fixing the coefficients of all pure $\phi$-interactions at a given order to zero fixes precisely the same number of independent coefficients as are present in the full potential at the same order, so no free parameters are left after eliminating the pure $\phi$-interactions at higher orders. The lowest orders are also fixed: quadratic order by Fierz-Pauli tuning, i.e. by requiring the linear theory to be ghost-free, and tadpole terms by requiring Minkowski to be a solution of the theory. So it is only at cubic and fourth order that non-vanishing total derivative combinations for $\phi$ exist and hence pure $\phi$-interactions can be set to vanish up to a total derivative and not identically, leaving the coefficient of the total derivative combination a free parameter. In four dimensions these two parameters are conventionally taken to be $c_3$ and $d_5$. In $D$ dimensions the number of free coefficients is $D-2$.  

Finally, we could have stopped the tuning of the potential at any point before raising the strong coupling scale to $\Lambda_3$. In fact, there is an infinite-parameter family of solutions with strong coupling scale $\Lambda_{\rm strong}$ arbitrarily close to $\Lambda_3$ from below. Such theories will however generically have operators such as $\left(\Box\hat\phi\right)^n$ for some large $n$ at the scale $\Lambda_{\rm strong}$, which will excite a ghost-like degree of freedom at that scale \cite{Creminelli:2005qk}. 

\subsection{The $\Lambda_3$ decoupling limit}

With the above dRGT-type tuning of the potential we have, by construction, eliminated all pure scalar interactions. The only other interaction terms below the scale $\Lambda_3$ are vector-scalar interactions of the form $ (\partial \hat{A}) (\partial^2 \hat{\phi})^n$. After the above tuning of the potential these terms are given by $\partial^{\mu} A^{\nu} \hat X_{\mu \nu}^{(n)}$, where 
\be
\hat X^{(n)}_{\mu\nu} =  \frac{1}{n+1} \frac{\delta}{\delta \hat\Pi_{\mu \nu}}\mathcal{L}_{n+1}(\hat\Pi),
\ee
and $\hat \Pi_{\mu\nu} \equiv \partial_\mu\partial_\nu\hat\phi$.
These satisfy  $\partial^{\mu} \hat X_{\mu \nu}^{(n)} = 0$, so the $ (\partial \hat{A}) (\partial^2 \hat{\phi})^n$ interactions also vanish with the same tuning of the potential. As a result, the strong coupling scale of these theories is $\Lambda_3$. 

In order to investigate the interactions at this scale, and hence those that are most important at low energies, we may now take the $\Lambda_3$ decoupling limit
\begin{align}
m &\rightarrow 0, &M_{P} &\rightarrow \infty, &\Lambda_3 &\text{ fixed},
\end{align}
where we have ignored the coupling to any external source \footnote{In this case we would typically take the decoupling limit with $T \to \infty, T/M_P \to \text{constant}$ in addition to the above, where $T$ is the trace of the stress-energy tensor of the source.}.
For dRGT, the $\Lambda_3$ decoupling limit action, without vector terms, is \cite{deRham:2010ik}
\begin{align} \label{lambda3decouplingdiag1} S = \int d^4x \; \frac{1}{2} \hat h_{\mu\nu}{\cal E}^{\mu\nu,\alpha\beta} \hat h_{\alpha\beta} -{1\over 2} \hat h^{\mu\nu}\left(-4\hat X_{\mu \nu}^{(1)}+{4(6c_3-1)\over \Lambda_3^3} \hat X^{(2)}_{\mu\nu}+{16(8d_5+c_3)\over \Lambda_3^6} \hat X^{(3)}_{\mu\nu}\right).
\end{align}
The other interactions contributing at the scale $\Lambda_3$ are scalar-vector terms of the form $(\partial \hat{A})^2 (\partial^2 \hat{\phi})^n$. These terms were explicitly calculated in \cite{Ondo:2013wka}, see also \cite{Gabadadze:2013ria}. From their expression, we can see that these terms cannot be removed with a special choice of parameters and so are always present in the decoupling limit. However, the vector $A$ always enters quadratically in this limit and this remains true when a coupling to matter is considered. As a result, the vector always appears at least linearly in the vector equations of motion. There is consequently a consistent classical solution for which the vector terms are set to zero \footnote{Note that, when computing scattering amplitudes, i.e.~the S-matrix, for the theory, contributions from the $(\partial \hat{A})^2 (\partial^2 \hat{\phi})^n$ interactions are unavoidable, however, even if the vectors had previously been set to zero. In this sense the vector-scalar interactions themselves get strongly coupled at the $\Lambda_3$ scale regardless of any tuning/solution one may impose on the vectors.}. We will restrict ourselves to these solutions for the remainder of the paper. 

\section{The minimal model and beyond the decoupling limit}
\label{sec:minimal}

In the previous section, we reviewed how raising the strong coupling scale of a massive gravity theory to $\Lambda_3$ uniquely singles out the two-parameter model of dRGT massive gravity. Using the improved expansion \eqref{eq:potentialexpansion2} over \eqref{eq:potentialexpansion} makes it clear that one cannot use the redundant coefficients of  \eqref{eq:potentialexpansion} to cancel physical interaction terms and further raise the strong coupling scale in this way. 

In this section we will discuss the so-called `minimal model' \cite{deRham:2010ik,Hassan:2011vm}, which is as a particular dRGT model corresponding to the parameter choice that makes the $\Lambda_3$ scalar-tensor interaction terms vanish. Since the decoupling limit interactions at $\Lambda_3$ vanish in this model, it is possible that, when the vector modes are set to zero, the strong coupling scale has been raised to some scale higher than $\Lambda_3$. Here we show explicitly that the resulting scale remains asymptotically close to $\Lambda_3$. 

This result is implicit in \cite{Renaux-Petel:2014pja}, which studied the Vainshtein mechanism in the minimal model and around particular configurations around massive sources of the minimal model (extending the minimal model Vainshtein analysis of \cite{Koyama:2011yg}). In that paper, it was pointed out that interactions suppressed by energy scales arbitrarily close to $\Lambda_3$ appear in generic time-dependent spherically symmetric solutions and generic non-spherically symmetric static solutions of the minimal model. We complement and extend this work by writing down such interactions at the level of the action without assuming any particular configurations, but working in full generality at the level of the action. Also note that in highly symmetric background solutions higher order interactions can of course vanish due to the symmetries imposed. For example, in spherically symmetric {\it and} static solutions the minimal model has no non-linear interactions up to the Planck scale \cite{Renaux-Petel:2014pja}. Here we will focus on the strong coupling scale of the theory as determined by the presence of interaction scales in the full action. 

\subsection{The minimal model}

From \eqref{lambda3decouplingdiag1}, it can be seen that the $\Lambda_3$ scalar and tensor interactions vanish for the particular parameter choices
\begin{equation}
\label{eq:minmodelparameters}
 c_3=1/6 \quad \text{and} \quad d_5=-1/48,
\end{equation}
which defines the minimal model as a unique dRGT theory \footnote{In higher dimensions, there is always a choice for the extra free parameters such that all decoupling limit interactions vanish.} (for fixed $m^2$ and $M_{P}$). Note that the kinetic mixing term 
\begin{equation}
\label{minmodint}
2\hat{h}^{\mu \nu} \hat X_{\mu \nu}^{(1)}
\end{equation}
is still present in the minimal model. This generates the scalar kinetic term after the linearised conformal field redefinition $\hat{h}_{\mu \nu} \rightarrow \hat{h}_{\mu \nu} + \hat{\phi} \eta_{\mu \nu}$, here expressed in terms of the canonically normalised fields, that demixes kinetic terms at quadratic order \footnote{Note that all interaction terms generated via applying this transformation to some original term inherit the same interaction scale as the original term. For example, the $\hat\phi^2 (\Box\hat \phi)^2$ interaction generated out of a ${\hat h}^2 (\Box\hat\phi)^2$ term will inherit the scale of the latter term.}. 
It is important that this kinetic term survives in the minimal model, since otherwise the presence of higher order interactions for $\pi$ would make the theory infinitely strongly coupled \cite{Hinterbichler:2011tt,Noller:2013yja}. 

To study the minimal model beyond the decoupling limit, we work with the re-summed version of the theory, which is given by \cite{Hassan:2011vm}
\be \label{resummedmm}
S_{min} = \frac{M_{\rm P}^2}{2} \int d^4 x \sqrt{-g} \left( R - 2m^2 \left(\Tr \sqrt{g^{-1} \eta }-3\right) \right).
\ee
Throughout this section we use matrix notation, so that, e.g. $g^{-1} \eta = g^{\mu \alpha} \eta_{\alpha \nu}$. 

\subsection{Interaction terms in the minimal model}

To study scalar-tensor interactions, we only need to consider the following term from \eqref{resummedmm}:
\be \label{minmodel}
\sqrt{-g} \; \Tr \sqrt{g^{-1} \eta },
\ee
since the other terms are invariant under diffeomorphisms and consequently will not contribute to interactions of the \St field $\hat\phi$. We can use \eqref{intscales} to determine the scale of any interaction terms after applying the \St trick, so we will ignore any coupling constants such as $m^2 M_P^2$ in front of terms for the time being. We now define the matrix ${\mathbb K}$ via
\be
{\mathbb K} \equiv -\delta + g^{-1}\eta, 
\ee
so that \eqref{minmodel} is given by 
\bea \label{eq:sqrtexpansion}
 \sqrt{-g} \; \Tr \sqrt{\delta + {\mathbb K} }  =  \sqrt{-g} \; \sum_{n=0}^{\infty} c_n  \Tr  {\mathbb K}^n , 
\eea
where 
\be
c_n = \frac{(-1)^n (2n)!}{(1-2n)(n!)^2 4^n}.
\ee
Now we introduce the scalar \St fields as in section \ref{Stuk} by replacing the background metric in \eqref{minmodel} as
\be \label{etastuck}
\eta \to  \eta - 2 \Pi + \Pi^2,
\ee
where we are ignoring vector modes as before. We also expand the inverse metric $g^{-1}$ in terms of the perturbation $h$ (which is defined by $g_{\mu\nu} = \eta_{\mu\nu} + h_{\mu\nu}$)
\begin{align}
g^{\mu \nu} & =  \eta^{\mu \nu} - h^{\mu \nu} + h^{\mu \sigma} h^{\; \;  \nu}_{\sigma} + {\cal O}(h^3)   \\
\sqrt{-g} &=  1 + \frac{1}{2} [h] - \frac{1}{4} [h^2] + \frac{1}{8}[h] [h] + {\cal O}(h^3)
\end{align}
where here, and in the rest of this section, $[\cdot]$ denotes a trace with the reference metric. Indices on $h$ and $\Pi$ are always raised and lowered with the reference metric $\eta$. We also write
\bea
{\mathbb K} = {\mathbb K}_{(0)} + {\mathbb K}_{(1)} + {\mathbb K}_{(2)} + {\cal O}(h^3) ,
\eea
where ${\mathbb K}_{(i)}$ is proportional to $h^i$, so that the first three orders are given by 
\bea
{\mathbb K}_{(0)} &=&  (-2\Pi + \Pi^2) \\
{\mathbb K}_{(1)} &=&  -h(\delta - 2\Pi + \Pi^2) \\
{\mathbb K}_{(2)} &=&  h^2(\delta - 2\Pi + \Pi^2).
\eea
Note that the higher order terms go up to arbitrary orders in $h$, but remain at most second order in $\Pi$ due to the form of the \St trick.

\subsection{Pure scalar interaction terms: scales below $\Lambda_3$}

The pure scalar terms are given by
\be \label{purescalar}
\sum_{n=1}^{\infty} c_n  \Tr  {\mathbb K}_{(0)}^n = -\left[ \Pi \right],
\ee
where we have used \eqref{eq:sqrtexpansion}. This is a total derivative, as expected. We see that in the minimal model the pure scalar terms vanish identically beyond first order. This is different to generic dRGT massive gravity models where pure scalar interactions up to and including fourth order in the fields vanish only up to total derivatives.

\subsection{Would-be decoupling limit interactions: the scale $\Lambda_3$}

Now we consider the scalar-tensor interactions at the scale $\Lambda_3$. These are the decoupling limit interactions, which should vanish by the definition of the minimal model. From \eqref{eq:sqrtexpansion}, the relevant terms are
\be \label{decint}
\sum_{n=1}^{\infty} n c_n \Tr \left( {\mathbb K}_{(1)}  {\mathbb K}_{(0)}^{n-1}  \right), \quad \text{and} \quad \frac{1}{2} \left[h\right] \sum_{n=1}^{\infty} c_n \Tr {\mathbb K}_{(0)}^{n} .
\ee
The first term comes from the following contributions:
\be \label{decint2}
\Tr \left( {\mathbb K}_{(0)}^{m_1} {\mathbb K}_{(1)}^{}  {\mathbb K}_{(0)}^{m_2}  \right) = \Tr \left({\mathbb K}_{(1)}  {\mathbb K}_{(0)}^{n-1}  \right),
\ee
where $m_1 + m_2 = n-1$ and we have used invariance of the trace under cyclical permutations \footnote{In fact, $\Tr \left( X Y Z \right) = \Tr \left( Y X Z \right)$, as long as $X,Y,Z$ are symmetric matrices. Note that this is not true for more than three matrices---we will come back to this below.}.  
We can expand and simplify the first term as follows:
\begin{align} \label{eq:decsimplify}
\sum_{n=1}^{\infty} n c_n \Tr \left( {\mathbb K}_{(1)}  {\mathbb K}_{(0)}^{n-1}  \right) & = -\sum_{n=0}^{\infty} \frac{c_n}{2}  
\left[ h (2\Pi+\Pi^2)^{n}   \right] \nn \\
& = -\frac{\left[h\right]}{2}+\frac{\left[h\Pi\right]}{2},
\end{align}
where we have used the identity 
\begin{equation} \label{identity1}
n c_n + (n+1)c_{n+1} = \frac{c_n}{2}
\end{equation}
 to get the first equality and the commutivativity of the trace and sum in \eqref{eq:sqrtexpansion} to get the second equality. The second term in \eqref{decint} has already been worked out in \eqref{purescalar}, so the total contribution is
\be
\frac{\left[h\Pi\right]}{2} - \frac{\left[h \right] \left[\Pi \right]}{2},
\ee
where we have drop the tadpole term since it is cancelled by a contribution from the determinant. 
 As expected, this is just the quadratic term (not suppressed by $\Lambda_3$) that survives the decoupling limit and generates the canonical scalar kinetic term after a field redefinition and canonical normalisation. No interactions survive at cubic order and higher in the $\Lambda_3$ decoupling limit in the minimal model.

\subsection{Higher order interaction terms: scales $\Lambda_{3-\epsilon}$ and above}

Since no scalar-tensor interactions (and hence no pure scalar interactions after demixing) survive at the scale $\Lambda_3$ in the minimal model, we now investigate what is the lowest scale associated with such interactions. This raised scale will be the new strong coupling scale of the theory in the absence of vector modes (see our discussion above) or, at the very least, will be the scale where the pure scalar interactions become strongly coupled. Establishing what this scale is requires us to go beyond the $\Lambda_3$ decoupling limit. Here we will show that, even for pure scalar modes, this strong coupling scale remains asymptotically close to $\Lambda_3$ in the minimal model.

We here explicitly compute $\{2,0,n-2\}$ terms (c.f. \eqref{curly}), which have interaction scales that asymptotically approach $\Lambda_3$ for large $n$, and show that these are indeed present in the action for arbitrarily large $n$.  The generic $\{2,0,n-2\}$ term with contraction between $h$ and $\Pi$ looks like
\be
\left[ h \Pi^k h \Pi^{n-2-k}\right].
\ee
These differ for different $k$, where $k$ runs from $1$ to $n-3$. If these are to vanish for a given $n$, they have to vanish for each $k$ independently, since we generally cannot permute inside the trace of a product of more than three matrices. 
To be specific, we consider terms of the form $ \left[h^2 \Pi^{n-2} \right]$ and $\left[h \Pi h \Pi^{n-3} \right]$ and show that these do not vanish at any order in the fields in the minimal model. \\

\noindent {\bf Interactions beyond the decoupling limit I:} From \eqref{eq:sqrtexpansion}, the terms contributing to $ \left[h^2 \Pi^{n-2} \right]$ interactions come from 
\be \label{h2terms}
 \Tr \left( {\mathbb K}_{(2)}  {\mathbb K}_{(0)}^{n-2}  \right), \quad\quad \text{and} \quad\quad \Tr \left({\mathbb K}_{(1)} {\mathbb K}_{(1)} {\mathbb K}_{(0)}^{n} \right).
\ee
By the same reasoning as in \eqref{eq:decsimplify}, the first terms in \eqref{h2terms} give in total
\be \label{firstsum}
\sum_{n=1}^{\infty} n c_n \Tr \left( {\mathbb K}_{(2)}  {\mathbb K}_{(0)}^{n-1} \right) =\frac{\left[ h^2 \right]}{2} - \frac{\left[ h^2 \Pi \right]}{2}.
\ee
 The total contribution from terms of the second type in \eqref{h2terms} turns out to be
\be \label{secondsum}
\sum_{n=2}^{\infty} c_n n \left[h^2(\delta-\Pi)^2(-2\Pi + \Pi^2)^{n-2}\right]
= \frac{1}{2} \left[ h^2 \left( -\delta +\frac{ \delta}{2 \delta - \Pi}\right)\right], 
\ee
where we have again used \eqref{identity1}. The right-hand side of \eqref{secondsum} can now be expanded as a Taylor series, giving, at cubic order in the fields and above, 
\be \label{fish1}
\frac{1}{4} \sum_{m=1}^{\infty} \frac{1}{2^{m}} \left[ h^2 \Pi^m \right],
\ee
which shows that scalar-tensor terms of the form $ \left[h^2 \Pi^{n-2} \right]$ survive to arbitrarily high orders.
\\

\noindent {\bf Interactions beyond the decoupling limit II:} Here we compute $\left[h \Pi h \Pi^{m-3}\right]$ interaction terms to show that these interactions beyond the $\Lambda_3$ decoupling limit also do not vanish, showing that there is nothing special about the $\left[h^2 \Pi^{m-2}\right]$ interactions considered above. Contributions now come from terms of the form
\be \label{aeqn1}
\text{Tr} \left({\mathbb K}_{(1)} {\mathbb K}_{(0)} {\mathbb K}_{(1)} {\mathbb K}_{(0)}^{n-3} \right) \quad\quad \text{and} \quad\quad \text{Tr} \left({\mathbb K}_{(1)} {\mathbb K}_{(1)} {\mathbb K}_{(0)}^{n-2} \right).
\ee
For sixth order in the fields and above these give 
\bea \label{fish2}
 \sum_{m=3}^{\infty} \frac{(m+1)}{2^{m+3}} \left[h \Pi h \Pi^m \right] - \sum_{m=3}^{\infty} \frac{1}{2^{m+1}} \left[ h \Pi h \Pi^m \right],
\eea
where the first (second) sum contains all contributions from the first (second) term in \eqref{aeqn1}. At quartic and quintic order the expressions for numerical coefficients of the low-order analogues of terms in \eqref{fish2} change due to additional redundancies between (and symmetries of) the terms in \eqref{aeqn1} at lower orders. 
Just as above we therefore find that beyond-the-decoupling-limit $\left[h \Pi h \Pi^{m-3}\right]$ terms remain to arbitrarily high orders in the minimal model. We emphasise that \eqref{fish1} and \eqref{fish2} are the only contributions of their respective forms and so can not be cancelled by other terms.  \\

\noindent {\bf Interaction scales:} Interaction terms of the form $\{2,0,n-2\}$, such as the two cases considered above, after canonically normalising the fields are suppressed by energy scales
\begin{equation} \label{bdlscales}
\Lambda_{\lambda_{(n)}}= m\left( \frac{M_P}{m} \right)^{1/ \lambda_{(n)}} , \ \ \ \lambda_{(n)} =3- {2\over n},
\end{equation}
which asymptotically approaches $\Lambda_3$ as $n \to \infty$. This explicitly confirms that interactions with energy scales arbitrarily close to $\Lambda_3$ remain in the minimal model beyond the decoupling limit, even though the $\Lambda_3$ decoupling limit interactions vanish. 

Note that the \St variables used here correspond to the helicity modes of the graviton as appropriate in the $\Lambda_3$ decoupling limit. In the minimal model, there is no decoupling limit that captures a finite set of interactions at a specific lowest energy scale and no such identification can be made, so that the fields used in the \St decomposition no longer describe helicity modes. However, we may still use this decomposition to infer the presence of particular interactions at a given scale \footnote{In other words, it is the physical interpretation of these interactions that is less straightforward now than it was before.}. \\

Also note that to demix the quadratic derivative terms we must redefine $h_{\mu \nu}$ with a linear conformal transformation. Such a field redefinition generates new terms that are suppressed by the same energy scale as the original term. For example, the $\Lambda_1=M_P$ term $-3[h^2 \Pi]/8$ produces the pure scalar term $-3\phi^2 [\Pi]/8$ and the cubic mixing term $-3\phi[h\Pi]/4$ after the linear conformal transformation, but these interactions are also suppressed by $M_P$ (once we have canonically normalised everything). It is interesting to further inspect the resultant pure scalar terms \footnote{We thank an anonymous referee for making this suggestion.} as these appear to be higher derivative $\sim \phi^a [\Pi^b]$ and hence may naively lead to the propagation of ghosts, even though we know that the theory is ghost free by previous results \cite{Hassan:2011hr, deRham:2011rn, Mirbabayi:2011aa, deRham:2011qq, Hassan:2011ea, Hassan:2012qv}. We can directly obtain the resummed pure scalar terms by substituting the conformally flat metric $g_{\mu \nu} = \eta_{\mu \nu}(1+ m^2 \phi)$ in \eqref{resummedmm} along with the \St replacement on the background metric \eqref{etastuck}. The pure scalar terms coming from the mass term (including the overall $\frac{M_P^2}{2} \sqrt{-g}$ factor) are
\begin{equation}
-m^2 M_P^2 \left( (4 - [\Pi]) (1+ m^2 \phi)^{3/2} - 3 (1+m^2 \phi)^2 \right).
\end{equation}
This shows that all the higher derivative pure scalar terms cancel, although the perturbative expansion in fields still has infinitely many terms. Expanding this perturbatively to lowest orders in the fields gives the expected result.
The surviving pure scalar terms are suppressed by energy scales greater than or equal to $M_P$, so we should also include the pure scalar terms coming from the Einstein-Hilbert kinetic term, which are suppressed by powers of $M_P$, but these will evidently not be higher derivative. This shows that after the field redefinition the leading interactions are still the scalar-tensor derivative mixing terms with scales close to $\Lambda_3$. These mixing terms have higher derivatives acting on $\phi$, so we may still worry about ghosts, but this is just because our variable choice is not useful for counting degrees of freedom beyond a decoupling limit. To see the absence of ghosts one should use different variables, e.g. this is shown using \St fields in \cite{deRham:2011rn, Hassan:2012qv} but without splitting the 4 independent diffeomorphism scalars into a Lorentz vector and scalar. 

\section{Conclusions} \label{sec-conc}

In this paper we showed that $\Lambda_3$ is the effective strong coupling scale around the Minkowski vacuum solution for {\it all} local and Lorentz-invariant massive gravity models with a Minkowski reference metric. In particular, we showed this for the minimal model of ghost-free massive gravity, where the strong coupling scale in the absence of vector modes (a classically consistent solution) is not obvious \footnote{This is in contrast to generic (dRGT) ghost-free massive gravity models, where this scale can just be read off from the decoupling limit interactions \cite{deRham:2010ik}. In fact, that $\Lambda_3$ is the maximal strong coupling scale for massive gravity in the presence of vector modes was already shown in \cite{ArkaniHamed:2002sp}. For the specific `minimal massive gravity' model (the unique dRGT theory where scalar-tensor interactions vanish at $\Lambda_3$) this was not known -- see related discussions in \cite{deRham:2010ik,Hinterbichler:2011tt}.}.
We explicitly showed the non-vanishing form of interaction terms remain in the action suppressed by scales larger than but arbitrarily close to $\Lambda_3$, i.e. the strong coupling scale effectively remains at $\Lambda_3$ \footnote{This is the same scale that is induced by vector modes, if they are present \cite{Ondo:2013wka}.}. This is consistent with the results of \cite{Renaux-Petel:2014pja}, who proved analogous results for the minimal model for particular field configurations around massive sources, whereas we worked directly at the level of the action without imposing any particular ansatz on the form of the metric as provided by the configurations considered in \cite{Renaux-Petel:2014pja}. 

In the process we also clarified the role of the redundant parameters that appear in the perturbative expansion of generic Lorentz-invariant massive gravity models. These parameters reflect redundancies that result from using the expansion \eqref{eq:potentialexpansion} beyond fourth order, so at this point one should use the simpler non-redundant expansion \eqref{eq:potentialexpansion2}. By explicitly constructing a non-redundant potential using Newton's identities as outlined by \cite{Hassan:2011vm}, we clarified why any massive gravity theory with a strong coupling scale of at least $\Lambda_3$ only has two free parameters, i.e. it is dRGT massive gravity. 

\begin{acknowledgments}
We would like to thank Clare Burrage, Pedro Ferreira, Kurt Hinterbichler, Macarena Lagos, Antonio Padilla, Sebastien Renaux-Petel, James Scargill and David Stefanyszyn for useful discussions. Some calculations were done using the computer algebra package xAct \cite{xAct}. JB is supported by the Rhodes Trust. JN acknowledges support from the Royal Commission for the Exhibition of 1851 and BIPAC.
\end{acknowledgments}

\bibliographystyle{JHEP}
\bibliography{paper-bib}

\end{document}